# Measurements of the ion concentrations and conductivity over the Arabian Sea during the ARMEX

**Devendraa Siingh, S.D. Pawar, V. Gopalakrishnan and A. K. Kamra**

**Indian Institute of Tropical Meteorology, Pune, India**

**Abstract**

Measurements of the small-, intermediate-, and large-ion concentrations and the atmospheric electric conductivity of both polarities have been made over the Arabian Sea on four cruises of ORV Sagarkanya during the Arabian Sea Monsoon Experiment (ARMEX)during the monsoon and pre-monsoon seasons of 2002 and 2003. Seasonally averaged values of the total as well as polar conductivity are much higher during the monsoon than pre-monsoon season. Surprisingly, however, the concentration of small ions are less and those of large and intermediate ions are more during the monsoon than pre-monsoon season. The diurnal variations observed during the pre-monsoon season show that the nighttime small ion concentrations are about an order of magnitude higher than their daytime values. On the contrary, the daytime concentrations of the intermediate and large ions are much higher than those of their nighttime values. No such diurnal variations in ion concentrations are observed in monsoon season. Also examined are the variations in ion concentrations of different categories with distance from the coastline in different seasons and the ion-concentration changes associated with the precipitation of various types that occurred over ORV Sagarkanya. It is sufficient to invoke the ion-aerosol attachment process to explain our pre-monsoon observations. However, the generation of highly charged large ions by the bubble-breaking process caused by the wave breaking due to strong southwesterly surface winds (10 – 12 m $s^{-1}$) over the Arabian Sea is postulated to explain the monsoon season observations.

## 1. Introduction

Cosmic rays are the main source of ionization over sea on a global scale. Production of ions and aerosols by breaking waves during the high wind speed periods and the raindrops splashing below raining clouds are other strong but local sources of ions at air-sea interface [*Blanchard*, 1963]. Normally, the high mobility small ions produced from these sources are mainly responsible for atmospheric electric conductivity over sea. However, these small ions soon get attached to aerosol particles and loose their mobility. Hoppel (1985) has solved the time-dependent ion-aerosol balance equations using the size-dependent ion-aerosol attachment coefficients and conclude that the assumption of charge equilibrium is valid for the typical ocean environment and is expected to be fairly good for environments, away from direct sources of particles. Since the intensity of cosmic rays does not significantly vary upto ~ $40^o$ of latitude, the atmospheric electric conductivity over sea is mainly determined by the aerosol particle concentration. Several measurements carried over different oceans of the world show that the atmospheric electric conductivity over ocean surface is inversely proportional to the aerosol concentrations. Consequently, the electric conductivity has often been suggested to act as an index of the background air pollution **[***Cobb and Wells,* 1970; *Morita*, 1971**;** *Misaki et al.*, 1972***;*** *Kamra and Deshpande*, 1995; *Deshpande and Kamra,* 1995**].** Measurements of the conductivity variation across the coastlines have also been used to study the extent of the transport of air pollution from land to ocean in the coastal regions [*Ruttenberg and Holzer,* 1955; *Misaki and Takeuti*, 1970; *Morita et al*., 1973; *Deshpande and Kamra*, 1995]. Kamra et al. [1997] reports a decrease in conductivity due to large values of relative humidity that exist in the region of Somali current. Based on the mobility spectrometry measurements of the atmospheric ions, Misaki et al. [1975] report that when the aerosols of land origin are dispersed over ocean, their size distribution is deformed in such a way that the center of gravity of the distribution is shifted towards smaller size with the increasing age of aerosols.

Large number of studies conducted in the last decade emphasize the role of atmospheric ions in the generation and growth of particles (e.g. *Arnold*, 1980; *Yu and Turco*, 2000; *Nadykto and Yu,* 2003; *Harrison and Carslaw,* 2003). Improvement in our measurement capabilities of aerosol size distributions down to 3 nm size over last decade has largely contributed to our understanding of the ion-aerosol interactions and the new particle formation in the atmosphere. Kulmala et al. [2004] has reviewed large number of such measurements made in different environments such as polluted continental boundary layer, marine boundary layer, upper troposphere, Arctic and Antarctic, boreal forests and coastal regions. These measurements suggest some aerosol formation mechanisms that can operate under specific meteorological conditions. Among a variety of different nucleation mechanisms proposed for particle formation, are the binary nucleation [*Kulmala and Laaksonen,* 1990], ternary nucleation [*Kulmala,et al.,* 2003], ion-induced [*Yue and Hamill,* 1979] and ion-mediated nucleation [*Yu and Turco,* 2000], mechanisms. The last two mechanisms involve strong ion-aerosol interactions in the atmosphere. *Yu* [2002], *Harrison and Carslaw* [2003], *Lovejoy et al.* [2004] discuss how the atmospheric ions and charged aerosols can cause particle formation in the atmosphere. *Tammet and Kulmala* [2005] propose a model which includes ion induced and homogeneous nucleation, depletion of ions and nanoparticles, the electric charges of particles to explain the atmospheric nucleation bursts such as those reported by *Horrak et al.* [1998].

Recently, Laakso et al. [2003, 2004a, b] have measured the ion production rate over boreal forests and discuss how these ions may be involved in the particle formation mechanisms.

Particle production in the aircraft or ships engine's exhaust is another source of their introduction in the atmosphere under some conditions (e.g. see Yu and Turco, 2000; Gopalakrishnan et al., 2005).

As a consequence of its utility to study the change in background air pollution, the atmospheric electric conductivity has been one of the frequently measured parameters over oceans. The conductivity is made up of different types of ions whose number concentrations and mobilities differ under different conditions. However, the relative contributions of various categories of ions to the electric conductivity under different meteorological conditions prevailing over ocean surface, has never been investigated. In this paper, we report our measurements of the small-, intermediate- and large-ion concentrations in addition to those of the conductivity of both polarities made over the Arabian Sea near to the west coast of India during the pre-monsoon and monsoon seasons in the Phase-I and II of the Arabian Sea Monsoon Experiment (ARMEX). To the author's knowledge, such simultaneous measurements of different categories of ions and conductivity over sea have never been made and are being reported here for the first time.

Measurements were made onboard ORV Sagarkanya during 2 cruises each of the ARMEX-2002 (Phase I) and ARMEX- 2003 (Phase II). The ship was kept at stationary position for 12-14 days during each cruise and the results obtained during these periods have been described elsewhere [*Pawar et al.***,** 2005]**.** Here, we present the diurnal and seasonal variations in ion concentrations and conductivity in the pre-monsoon and monsoon seasons and discuss the changes in these parameters associated with some particular meteorological conditions.

## 2. Instrumentation

Measurements of the small-, intermediate-, and large-ion concentrations are made with an ion-counter consisting of three separate condensers. Table 1 shows the mobility and particle size ranges covered by the small-, intermediate-, and large-ion condensers**.** The ion-counter is similar to that described by Dhanorkar and Kamra [1991]. It need be mentioned here that transfer functions for three ion groups will determine the precise discrimination of mobility groups in three ion counters. Each condenser consists of two-coaxial cylinders with the outer cylinder raised to a higher potential with respect to inner one, and is shielded with another earthed coaxial cylinder around it. Air is drawn through each of the three condensers with a common fan fixed in a cylinder fitted at one end of the condensers. Table *2* shows dimensions and potentials applied to the three condensers. The lowest concentration of ions that can be reliably measured by the small-, intermediate-, and large-ion condensers are 2, 9, and 54 ions $m^{-3}$, respectively. Atmospheric electric conductivity of both polarities is measured with a Gerdien's apparatus consisting of two identical coaxial condensers each with a critical mobility of $3.6 \times 10^{-4}\ m^2\ V^{-1} s^{-1}$. Details of the apparatus are described by Dhanorkar and Kamra [1992].

Both instruments, ion counter and Gerdien's apparatus were installed on the balloon launching platform of the ship with their intakes perpendicular to the ships direction of motion and ~ 9 m above mean sea level. The balloon launching platform is 34 m away from the ship's chimney and 15 m below the chimney's exhaust. So the chances of the exhaust polluting the measurement site and contaminating the data are very small unless there is strong downward motion in the vicinity of the chimney. However, the data obtained during the periods when the measurement site was visually observed to be polluted, which happened very rarely or when the measured variables experienced some adrupt and abnormal changes are not included in our analysis. Stability of the instruments was maintained by cleaning the insulators and checked by checking the zero-shift of each output at least twice daily and correcting it, if necessary. No significant zero-shift was observed except during the periods of heavy rain Therefore, observations taken during periods of heavy rain were not included in our analysis. After each incident of precipitation the apparatus was cleaned, dried and put in operation after checking its zero.

**Cruise and Weather**

Figure 1 shows the four cruise routes – one in the pre-monsoon season (March 14 to April 10, 2003), one around the monsoon-onset period (May 15 to June 19, 2003) and two in monsoon season (June 21 to July 15 and July 17 to August 16, 2002). The stationary position during each cruise is shown by a circle. Continuous observations of all parameters, except rainfall, are made throughout the cruises. However, observations could not be made during some time- periods in the beginning and end of the cruises because of the lack of sufficient time required to install the equipment before the cruise starts or the time required to pack-up the instruments at the end of the cruise. Moreover, there are large data-gaps due to the time required for the frequent cleaning and maintenance of the equipment in the marine atmosphere, rain spells and other logistic and technical difficulties. As a result of these data-gaps and rejection of some data due to possible contamination as discussed above, an average of only 74%, 80%, 70% and 55% of data recorded in the observational periods of the pre-monsoon, monsoon-onset, monsoon I and monsoon II cruises, respectively, were available for analysis.

The southwest monsoon extending from June to September, each year, is the dominant feature of this region when a seasonal wind flows with consistency and regularity from the southwest direction . Towards the end of May, a southwesterly wind from Somalia coast spreads northwards over the Arabian Sea, the Bay of Bengal and the Indian sub-continent. The onset of southwesterly winds over the west coast of India is often sudden. Generally, it is referred to as the bursts of the monsoon over India. Strong winds continue to blow most of the time during the southwest monsoon period and exhibit periodic movements to the north and south of the mean location of the tropical easterly jet. Normally 75% of annual rainfall over India is received during the southwest monsoon season. However, the rainfall during the monsoon has short period rainfall fluctuations and is characterized by heavy rain followed by lean periods known as break monsoon. The southwest monsoon withdraws from this region at the end of September. Winds undergo a reversal in direction and weak northeasterly winds flow in winter monsoon season.

The NCEP average winds for the period of each cruise are shown in Figure 2. The diagrams clearly show the systematic transition from weak northerly winds during the pre-monsoon cruise to strong southwesterlies during the monsoon season. The difference in ion concentrations during different airflows and the conductivity caused by them are discussed below.

## Observations

### Seasonal Average Ion Concentrations

Table 3 shows the average values alongwith their standard deviations of the small-, intermediate-, and large-ion concentrations *of the positive polarity* and the positive, negative and total conductivity at the four stationary positions. Large variability of each parameter within an individual cruise especially during the monsoon season, is noticeable. Total conductivity as well as the conductivity of each polarity are much higher during the monsoon than the pre-monsoon periods. Surprisingly, the concentration of small ions which contributes most to the conductivity, is less during monsoon season than in the pre-monsoon season. However, the concentrations of the large and intermediate ions are much more during the monsoon periods as compared to the pre-monsoon periods.

Table 4 shows the average values of the parameters for the whole period of each cruise instead of only for the stationary period. Although, the magnitudes of individual corresponding values in Tables *3* and *4* are little different from each other, the trends in variation of average values in different cruises (or seasons) are similar. The observation indicates that the time-averaged values of ion concentrations along a cruise in this area fairly well represent the temporal/ seasonal changes of the various ion-category concentrations.

**Diurnal Variations**

All the three categories of ions and both polar conductivities show diurnal variations in the pre-monsoon period. Figure 3 shows these variations on a typical fair-weather day on May 19, 2003. The nighttime small ion concentrations (from 1800 – 2000 to 0600 – 0900 hours) are about an order of magnitude higher than their daytime values. On the contrary, the daytime intermediate and large ion concentrations are much higher than their nighttime values. Moreover, contrary to the small-, and large-ion concentrations which experience large fluctuations during nighttime as compared to daytime values, the intermediate ion concentrations remain comparatively steady during daytime but experience large fluctuations during nighttime. The corresponding polar conductivities are little higher but much more fluctuating during daytime than during nighttime. During the pre-monsoon season, the variations in small ion concentrations are generally found to be inverse of those in large ion concentrations.

The situation changes much with the onset of monsoon season. The systematic diurnal variations observed in ion concentrations and conductivity in the pre-monsoon season, disappear during the monsoon-onset and monsoon periods. Values of ion concentrations and conductivity arbitrarily change over large ranges and do not show any regular variation from day to day. Moreover, the concentration variations observed in one ion-category become independent of variations in other ion-category.

**Variations of Ion Concentrations with Distance from the Coastline**

The land-to-sea transport of aerosols, as studied by *Ruttenberg and Holzer* [1955], *Misaki and Takeuti* [1970], Morita et al. [1973], *Deshpande and Kamra* [1995], is one prominent process in determining the concentration of ions and aerosols across coastlines. Several other sources and sinks can operate and influence their concentrations near coastlines. For example, the processes of nucleation, bubble bursting and raindrop splashing can introduce new ions and aerosols into the atmosphere. On the other hand, the processes of surface deposition, in-cloud and below cloud scavenging can remove them from the atmosphere. Fog also significantly influences their concentrations (e.g. Deshpande and Kamra, 2004). Air mass history at the place of measurements is another factor that strongly influences the ion and aerosols concentrations. Our present measurements are not sufficient to assess the contributions of different processes.

Unfortunately, our observations could not be made in the first 50 – 150 km from the coastline because the time available between the loading of the instruments and scientists onboard the ship and the start of the cruise was not sufficient to install the equipments. However, the variations in ion concentrations and conductivity observed from ~ 100 km to 400- 500 km from the coastline on some legs of the cruise showed following features in different seasons.

In the pre-monsoon period in Figure 4c, the trends in variations of the small and intermediate ion concentrations with distance are almost opposite to that of large ions. The small- and intermediate-ion concentrations increase upto ~ 250 km, decrease by almost two orders of magnitude at ~ 370 km and then again start increasing beyond 370 km. On the contrary, the large ion concentrations decrease upto 300 km, sharply increase by almost two orders of magnitude at ~ 370 km and then again start decreasing beyond ~ 370 km. The conductivity shows a general decrease upto ~ 320 km except some higher values around ~ 250 km and then settles down to comparatively constant values beyond ~ 320 km (Fig. 5).

Ion concentrations of all categories show perhaps most unique distributions with distance in the monsoon-onset period in Figure 4d from all other seasons. Concentrations of all category of ions remain almost constant for long distances, but show some rather abrupt changes. Most noteworthy are the sharp decrease at ~ 200 km and rather gradual increases in the 50 – 100 and 300 – 400 km regions in the large ion concentration. Corresponding values of conductivity in Fig. 5d also show almost uniform values upto ~ 300 km and then some increasing trend beyond 300 km.

Except for some very high values of large ion concentrations at ~ 200 km, the intermediate and large ion concentrations remain nearly constant from 150 to 500 kms in the monsoon-I period in Fig. 4a. The small ion concentration during this period, however, fluctuate over a large range. The conductivity during this period, sharply decreases to low values at some locations but soon recovers to values somewhat higher than during the monsoon onset season.

**Effect of Precipitation**

Observations of only total rainfall during 24 hours period were made during the cruise. Rain intensity was not measured. However, some incidents of rain were manually recorded. Therefore, the details of the time, period or intensity of rain wherever available in the following cases, need be taken as approximate only. We present below, three cases in which rainfall produced three distinct types of changes in ion concentrations of different categories :

**Case I:** With the arrival of the southwest monsoon a rainfall of 10.4 mm was recorded on the first day, on June 7, 2003 on board of Sagarkanya; the whole of rainfall falling in the morning hours. With the onset of showers, the meteorological records made at ORV Sagarkanya (Figure 6) showed a fall of ~ 3°C in atmospheric temperature, a rise of ~ 17% in relative humidity and a sudden burst of southwesterlies of upto ~ 11 m $s^{-1}$. Concentrations of the small and intermediate ions decreased as shown in Figure 6, most probably because of the scavenging of ions and aerosols by raindrops. After the showers, the concentrations of small and large ions, however, started increasing because of the cessation of the scavenging process, but the intermediate ion concentration continued to fall, perhaps due to no fresh formation of intermediate ions under the post-shower meteorological conditions. Magnitude of negative conductivity initially decreases at 0200 hours but then recovered to its original level at ~ 0500 hours. Nearly constant values of conductivity observed around 0800 hours, inspite of decrease in small and intermediate ion concentrations are most likely due to introduction of higher mobility ions of either category. Almost noise-free values of conductivity and intermediate ion concentrations before 0200 hours reflect that the changes occurring in these parameters are less than the sensitivity of these channels.

**Case II:** Drizzle type of rain produced different types of changes in ion concentrations. On July 23, 2002, drizzle started at ~ 0800 hours and continued for ~ 30 minutes. Thereafter, intermittent periods of light drizzle and mild sunlight continued upto 1225 hours when moderate drizzle occurred continuously from 1225 to 1320 hours. Strong winds of 9 – 12 m $s^{-1}$ from southwest continued for the whole day. As shown in Figure 7, ion concentrations in any category did not show any sharp or adrupt changes at 0800 hours. However, the large and intermediate ion concentrations started to gradually decrease with the start of the drizzle period but started increasing after 1330 – 1400 hours. On the contrary, the small ion concentrations showed an opposite trend i.e. an increase with the start of drizzle and decrease after the drizzle. (Figure 7). The small ion signal was out of the range for some periods between 1230 and 1600 hours. The electrical conductivity of both polarities somewhat increased during the drizzle but decreased soon after its ceasation. Comparatively smaller changes in conductivity during diurnal cycle in view of about an order of magnitude change in the small ion concentration indicate introduction of lower-mobility small ions and/or reduction in mobility of the existing small ions during the drizzle period. It may also be partially ascribed to the reduced values of intermediate/large ion concentrations during the drizzle period. The above results of ion concentrations can be interpreted if one considers the following two mechanisms. Firstly, the collection efficiency of drizzle size drops for small ions is much smaller than those for the intermediate or large ions. Secondly, the film-drops produced by splashing of drizzle drops on sea surface introduce some highly charged positive ions of larger size into the atmosphere (*Chapman,* 1938; *Gathman and Hoppel,* 1970a; *Pawar et al.,* 2005). In addition, processes such as the evaporation of ions from the aerosol particles and/or Rayleigh explosions of drying rain droplets can significantly contribute to the ion-aerosol budget in such environmental conditions.

**Case III:** On June 5, 2003, at ~ 0600 hours the atmospheric temperature suddenly dropped by ~ $3^{o}$ C, relative humidity increased by 18% and weak southwesterlies of ~ 2 m $s^{-1}$ changed to strong southerlies of ~ 12 m $s^{-1}$ (Figure 8). Light rain was recorded at ~ 0730 hours. However, its time of ceasation could not be recorded. The sharp increase at 0600 hours followed by a slow decrease in large ion concentration in Figure 8 as the winds slow down indicates introduction of wind-produced large ions in the atmosphere. These changes may also be due to change in air mass as the wind direction change from northwest to southwest and atmospheric temperature drops by 3° C (Figure 8). The small ion concentration continued to increase upto ~ 1000 hours when it decreased to its original level in ~ 20 minutes. The intermediate ion concentrations remained almost constant during this period.

**Discussion**

Our results can be interpreted in terms of the meteorological features and their changes in different seasons of the region. In the pre-monsoon period, the surface winds over the Arabian Sea and the northern Indian Ocean are weak, turbulent and frequently change their directions. So, the aerosols of land origin can be transported to the measurement locations over the sea ***with*** winds and/or by eddy diffusion**.** Under such conditions, the values of the ion concentrations in different categories and conductivity will be determined by the equilibrium reached by the ion-aerosol attachment process. Comparatively higher small-ion concentrations in the pre-monsoon than monsoon periods are most likely due to the transport of ions and radon from land, especially from the coastal regions where the sand is known to have very high radioactivity. The life-time of small ions is only of the order of a few minutes and thus these are not likely to be transported far away from the coastline before they get attached to aerosol particles. However, the presence of radon can cause in-situ production of small ions. In case of the land-to-sea transportation, therefore, concentrations of small ions and neutral aerosol particles should decrease but of charged aerosol particles should increase with distance from the coastline. Our results in Figure 4 c support such transportation. Further, the intermediate ion-concentration in our observations are much higher than those observed by *Horrak et al*. [1998] at Tahkuse Observatory in Estonia. It is partly because of much wider range of the intermediate-ion mobility in our ion-counter. Another, and probably more dominant reason may be higher rate of intermediate ion production in tropical areas where atmospheric temperatures and solar radiation are much higher than at Tahkuse. Availability of plenty of radon-produced small-ions and sulphate components being transported from land to sea in conditions of relatively low relative humidity in the pre-monsoon season will also enhance the formation of intermediate ions. Over and above these high concentrations no bursts of intermediate ions produced by nucleation of intermediate ions as suggested by *Kulmala et al.* (2004) and *de Reus et al.*, [1998] and are observed by *Horrak et al.,* [1998], are observed in our measurements. Unusually large intermediate-ion concentrations during the June 30 to July 12, 2002 period may be due to anomalous behaviour of monsoon in 2002 when the month of July, normally the wettest month in this region, was very dry, with clear skies, almost no precipitation and comparatively weaker southwesterly flow.

Onset of the southwest monsoon season follows establishment of cross-equatorial flow between 40 and 60$^{o}$ latitudes and strengthening of the southwesterly surface winds over the Arabian Sea and the north Indian Ocean. An area of strong winds starts extending eastwards from the Somali coast. The area of low winds prevailing over the central Arabian Sea during the pre-monsoon season starts getting smaller and slowly almost disappears. The onset of southwest monsoon is characterized by high surface winds which cause breaking waves at the sea surface. Breaking bubbles at the sea-air interface eject highly charged water droplets into the atmosphere [*Blanchard*, 1963, 66]. These droplets soon evaporate in the unsaturated atmosphere and leave behind highly charged residue particles. Large space charge densities have been observed to originate in the surf zones and move along with the wind [*Blanchard,* 1966; *Gathman and Hoppel*, 1970a, b; *Gathman and Trent*, 1968]. These field investigators indicate that the particles produced by the bubble breaking process are highly charged and have large mobility.

Laboratory experiments of *Chapman* [1938], *Woolf et al*. [1987], support such conclusion. These highly charged residue particles not only increase large ion concentrations but can also contribute to the atmospheric electric conductivity [*Pawar et al*., 2005]. Occurrence of comparatively very large concentrations of large ions in the monsoon months in Tables 3 and 4 and their good correlation with wind speed supports such process of the large ion generation. Similar trends in the average concentration values of different categories of ions in Tables 3 and 4 in different seasons support the fact that the temporal variations in the area of measurements can be reasonably well represented by the spatial averages for the whole cruise.

Figure 9 shows four typical 5-day back trajectories drawn from NOAA-HYSPLIT model on four days during the measurement periods. The direction of transportation of the airmass reaching the measuring site slowly shifts from North (almost along the western coastline of India) in the pre-monsoon season to the west and then to southwest in the monsoon season. With the setting-in of the cross-equatorial flow at 40 to 60$^{o}$E latitudes at the time of monsoon onset, the airmass being brought to the measuring site is from the southern hemisphere during the monsoon period. Southwesterly winds in the monsoon season push the aerosols and other pollutants, including Radon, of the land-origin and limit their extension over sea. Also, the higher southwesterly winds generate the highly charged ions at the air-sea interface and transport them towards the western coastline of India. Scatter plots of large ion concentration vs wind speed plotted for our data of monsoon-onset period (see Figure 5 in Pawar et al., 2005) shows a good correlation between the two parameters and stress the importance of wind-generated ions. Thus, two sources of large ions – first, the ion-aerosol attachment process prominent at land surface, and second, the bubble breaking process prominent in high wind speed regions at sea surface act together in monsoon season to determine the net large ion concentration near the coastline.

Higher concentrations of large ions during daytime and the opposite trends in variations in the small- and large -ion concentrations reflect the dominance of the first process in pre-monsoon season. On the other hand, the observations of very large ion concentrations and almost uncorrelated variations in ion concentrations of different categories indicate dominance of the second process in monsoon season.

Falling precipitation particles below clouds scavenge ions and aerosols in the subcloud layer. Reductions in ion concentrations of all categories following rain showers and subsequent recoveries of small and large ion concentration after rain on June 7, 2003, well demonstrate such removal of ions by scavenging process. The decrease in the intermediate ion concentration even after the showers is perhaps due to ceasation of intermediate ion formation process in the meteorological conditions that prevail after showers. Although our observations on July 23, 2003 (Fig. 7) also show reduction in large and small ion concentrations following drizzle, the increase in small ion concentration during the drizzle period and their decrease after the drizzle, perhaps, indicates generation of small ions due to drizzle drops splashing on the sea surface. Processes of the evaporation of falling charged drops and/or Rayleigh bursting can also contribute to it.

**Acknowledgements**

We thank organizers of the ARMEX for our participation in the experiment. We are grateful to Dr G S Bhat for providing the meteorological data collected onboard Sagarkanya. The authors gratefully acknowledge the NOAA Air Research Laboratory (ARL) for the provision of the HYSPLIT transport and dispersion model and/or READY website (http://www.arl.noaa.gov/ready.html) used in this publication

**Table 1: The mobility and size ranges of ions**

| Category | Small ions | Intermediate ions | Large Ions |
|---|---|---|---|
| **Mobility range ($m^2 V^{-1} m^{-1}$)** | **> 0.77 x $10^{-4}$** | **1.21 x $10^{-6}$ - 0.77 x $10^{-4}$** | **0.97 x $10^{-8}$ -1.21 x $10^{-6}$** |
| **Diameter range (nm)** | **< 1.45** | **1.45 – 12.68** | **12.68 - ~130** |

**Table 2:** Dimensions and other parameters of three condensers of the ion-counter.

| Dimensions/constants | Small-ion Condenser | Intermediate-ion Condenser | Large -ion Condenser |
|---|---|---|---|
| Length of the outer electrode (m) | 0.4 | 0.8 | 1.2 |
| Length of the inner electrode (m) | 0.2 | 0.5 | 1.0 |
| Diameter of the outer electrode (m) | 0.098 | 0.06 | 0.038 |
| Diameter of the inner electrode (m) | 0.076 | 0.037 | .022 |
| Potential applied (V) | 15 | 100 | 600 |
| Critical mobility ($m^2 V^{-1} m^{-1}$) | $0.766 \times 10^{-4}$ | $1.2 \times 10^{-6}$ | $0.97 \times 10^{-8}$ |
| Flow rate ($l\ s^{-1}$) | 8.6 | 1.8 | 0.29 |

**Table 3:** Periods and positions of ship at stationary positions and average values of small-, intermediate-, and large- ion concentrations ($N_s$, $N_I$ and $N_L$ respectively) of positive polarity and polar ($\Lambda_+$ and $\Lambda_-$) and total ($\Lambda$) conductivity during four cruises in the premonsoon, monsoon-onset and monsoon seasons. Values in parenthesis show their standard deviations.

| Period | Location of ship (Stationary period) | $N_S$ x $10^{-6}$ ($m^{-3}$) | $N_I$ x $10^{-6}$ ($m^{-3}$) | $N_L$ x $10^{-6}$ ($m^{-3}$) | $\Lambda_+$ x $10^{-14}$ ($Sm^{-1}$) | $\Lambda_-$ x $10^{-14}$ ($Sm^{-1}$) | $\Lambda$ x $10^{-14}$ ($Sm^{-1}$) |
|---|---|---|---|---|---|---|---|
| June 30- July 12, 2002 (Monsoon period -I) | $16.9^0$ N, $71.2^0$ E | 842 (±479) | 3060 (±849) | 6859 (±1980) | 0.89 (±0.34) | 0.83 (±0.17) | 1.72 |
| July 23– August 4, 2002 Monsoon Period -II) | $15.4^0$ N, $72.2^0$ E | 373 (±249) | 1081 (±599) | 11401 (±7267) | 0.83 (±0.68) | 0.65 (±0.27) | 1.48 |
| March 25- April 5, 2003 (Pre-monsoon period) | $9.1^0$ N, $74.5^0$ E | 895 (±557) | 1116 (±1241) | 3571 (±1549) | 0.47 (±0.12) | 0.54 (±0.23) | 1.01 |
| May 23- June 7, 2003 (Monsoon onset period ) | $9.1^0$ N, $74.5^0$ E | 1515 (±741) | 784 (±357) | 6090 (±3510) | 0.55 (±0.17) | 0.51 (±0.20) | 1.06 |

**Table 4:** Periods and area of ship and average values of small-, intermediate-, and large- ion concentrations ($N_s$, $N_I$ and $N_L$ respectively), of positive polarity and polar ($\Lambda_+$ and $\Lambda_-$) and total ($\Lambda$) conductivity during four cruises. Values in parenthesis show their standard deviations in the pre-monsoon, monsoon-onset and monsoon seasons.

| Period | Area | $N_S$ x $10^{-6}$ ($m^{-3}$) | $N_I$ x $10^{-6}$ ($m^{-3}$) | $N_L$ x $10^{-6}$ ($m^{-3}$) | $\Lambda_+$ x $10^{-14}$ (S/m) | $\Lambda_-$ x $10^{-14}$ (S/m) | $\Lambda$ x $10^{-14}$ (S/m) |
|---|---|---|---|---|---|---|---|
| June 21- July 15, 2002 (Monsoon period -I) | $15^0 – 17^0$ N, 69.5–$73.3^0$ E | 935 (±637) | 2424 (±2804) | 5875 (±5397) | 0.70 (±0.40) | 0.75 (±0.24) | 1.45 |
| July 17– August 16, 2002 (Monsoon period -II) | 14.4 -$8.3^0$ N, 71.5-$72.7^0$ E | 371 (±240) | 1654 (±1311) | 11640 (±9052) | 0.99 (±0.78) | 0.83 (±0.78) | 1.82 |
| March 14 - April 10, 2003 (Pre-monsoon period) | 8.3-$16.9^0$ N, 76.3-$72.5^0$ E | 1106 (±883) | 1188 (±955) | 4803 (±3436) | 0.53 (±0.15) | 0.62 (±0.25) | 1.16 |
| May 15- June 19, 2003 (Monsoon-onset period) | 7.5-$12.5^0$ N, 71.4-$76.4^0$ E | 1308 (±662) | 870 (±405) | 5528 (±3166) | 0.50 (±0.17) | 0.50 (±0.21) | 1.00 |

**Legends**

Fig. 1: Cruise track of cruises (a) during monsoon period and (b) during pre-monsoon and around monsoon-onset period.

Fig. 2: Average NECP derived winds for cruise periods. (a) June, 2002 to July, 2002; (b) July, 2002 to August, 2002; (c) March 15, 2003 to April 9, 2003; and (d) May 16, 2003 to June 19, 2003.

Fig. 3: Typical diurnal variation of small-, intermediate- and large-ions concentration and polar conductivities observed on a fair weather day.

Fig. 4: Variation of small-, intermediate- and large-ions concentration with distance from coast during (a) June 21, 2002 to July 15, 2002; (b) July 17, 2002 to August 16, 2002; (c) March 14, 2003 to April 10, 2003; and (d) May 15, 2003 to June 19, 2003.

Fig. 5: Variation of total conductivity with distance from coast (a) June 21, 2002 to July 15, 2002; (b) July 17, 2002 to August 16, 2002; (c) March 14, 2003 to April 10, 2003; and (d) May 15, 2003 to June 19, 2003.

Fig. 6: Variation with local time of temperature, relative humidity, pressure, wind speed and wind direction, small-, intermediate- and large-ion concentrations and polar conductivities observed onboard ORV sagarkanya on 7 June, 2003.

Fig. 7: Same as Fig. 6 but for July 23, 2002.

Fig. 8: Same as Fig. 6 but for June 5, 2002.

Fig. 9: 5-days NOAA-HYSPLIT back trajectories on four different days during the measurement periods.

Fig. 9

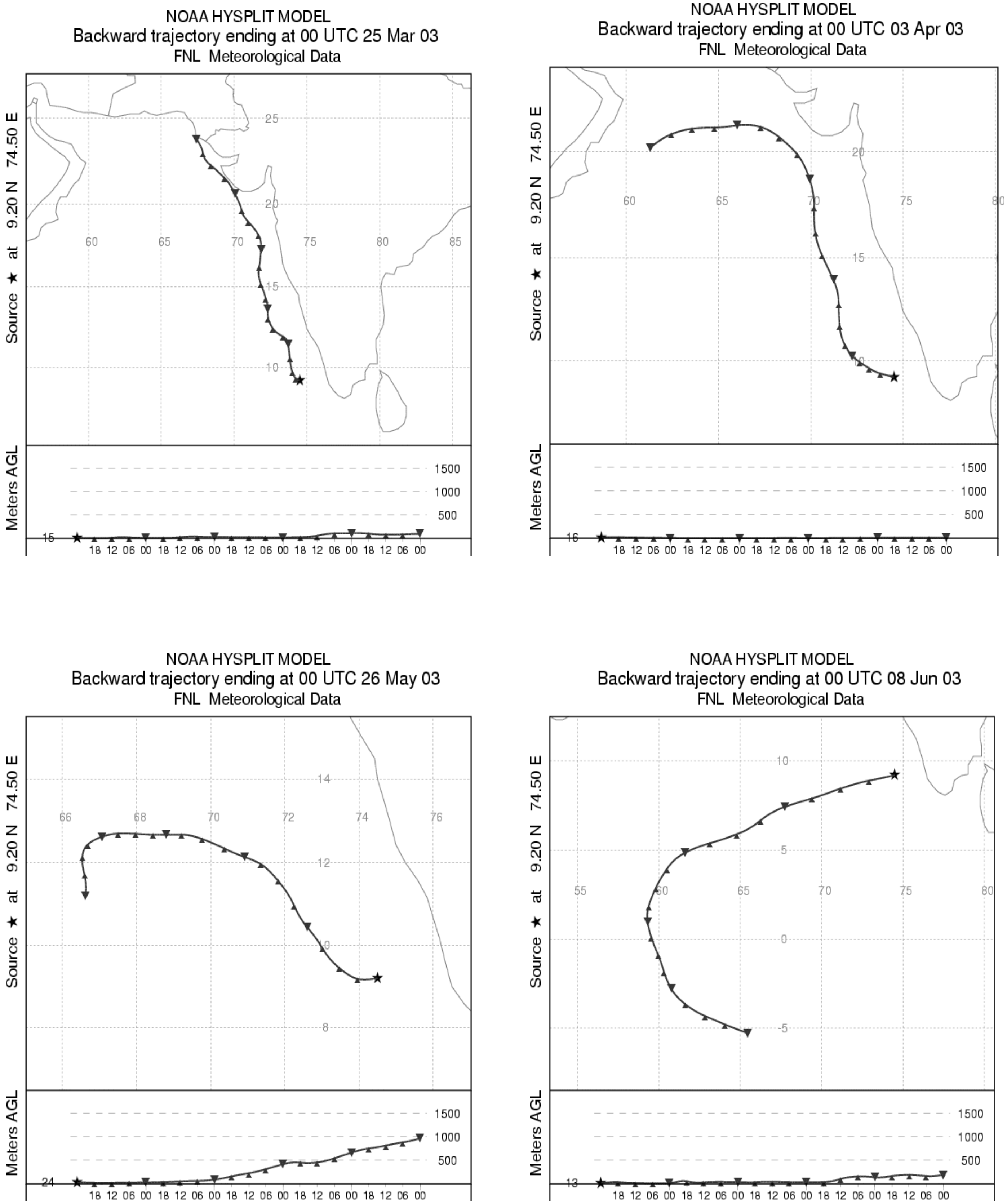
NOAA HYSPLIT MODEL
Backward trajectory ending at 00 UTC 25 Mar 03
FNL Meteorological Data
Source ★ at 9.20 N 74.50 E
Meters AGL
NOAA HYSPLIT MODEL
Backward trajectory ending at 00 UTC 03 Apr 03
FNL Meteorological Data
Source ★ at 9.20 N 74.50 E
Meters AGL
NOAA HYSPLIT MODEL
Backward trajectory ending at 00 UTC 26 May 03
FNL Meteorological Data
Source ★ at 9.20 N 74.50 E
Meters AGL
NOAA HYSPLIT MODEL
Backward trajectory ending at 00 UTC 08 Jun 03
FNL Meteorological Data
Source ★ at 9.20 N 74.50 E
Meters AGL
1500
1000
500

Fig. 2

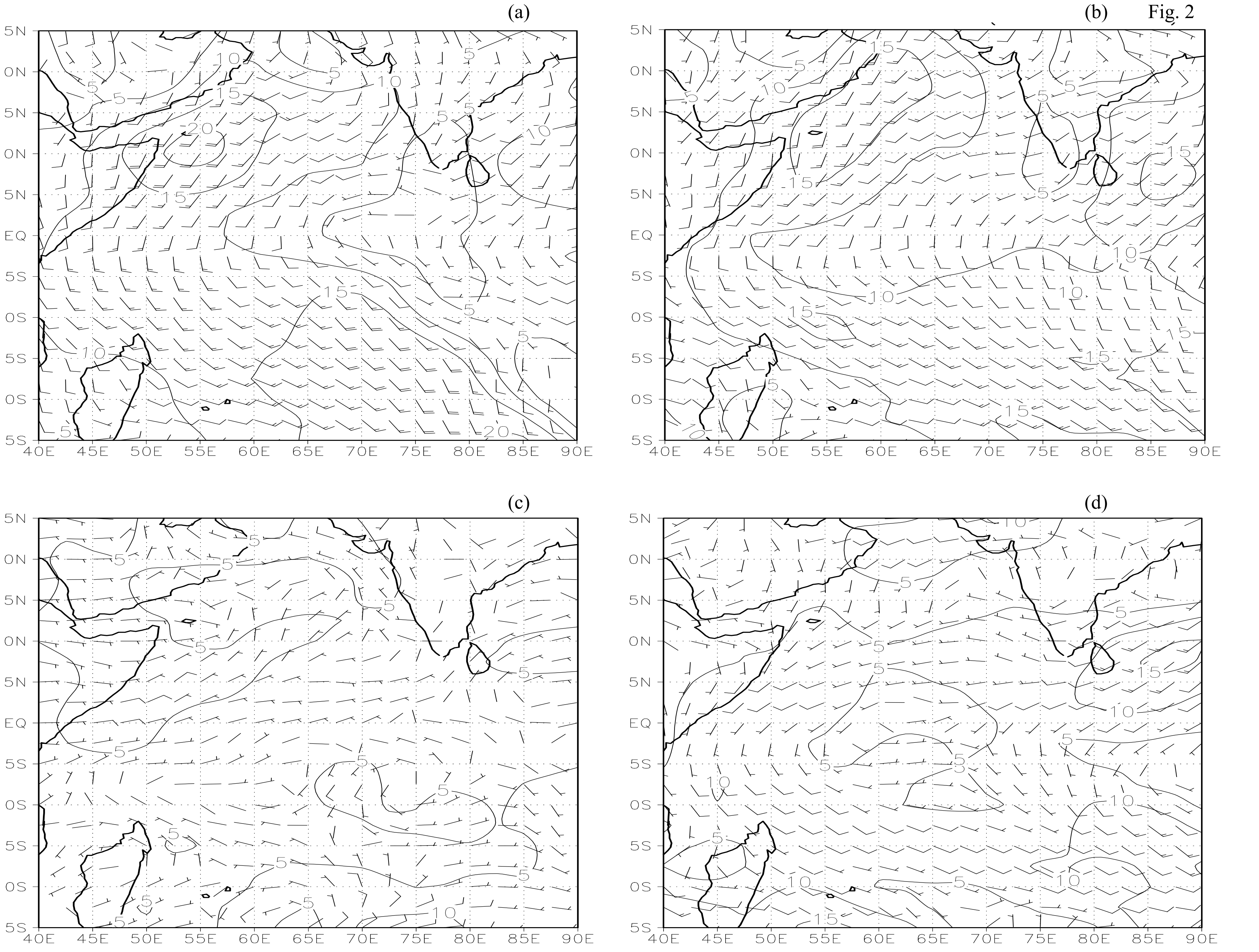

Fig. 1

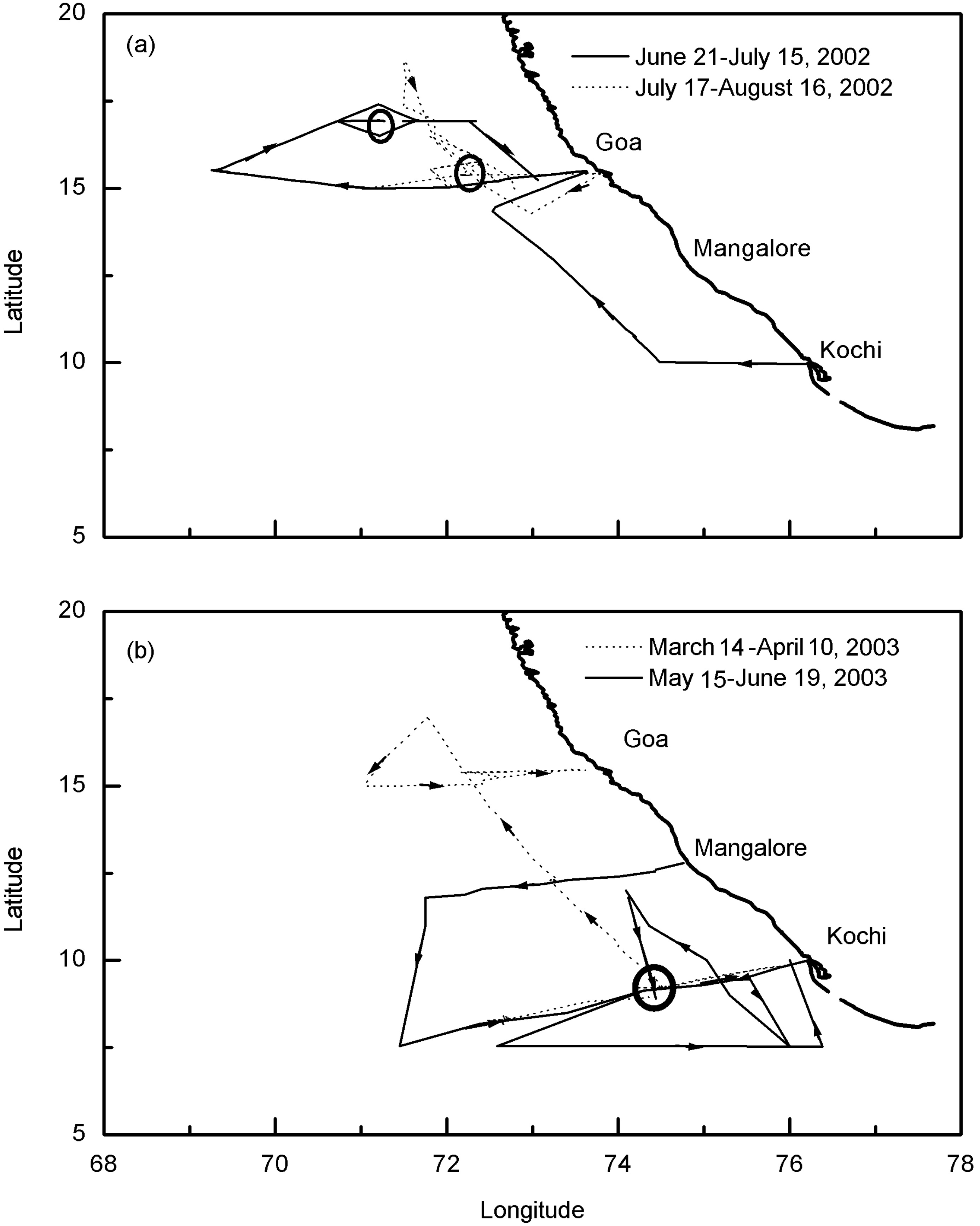

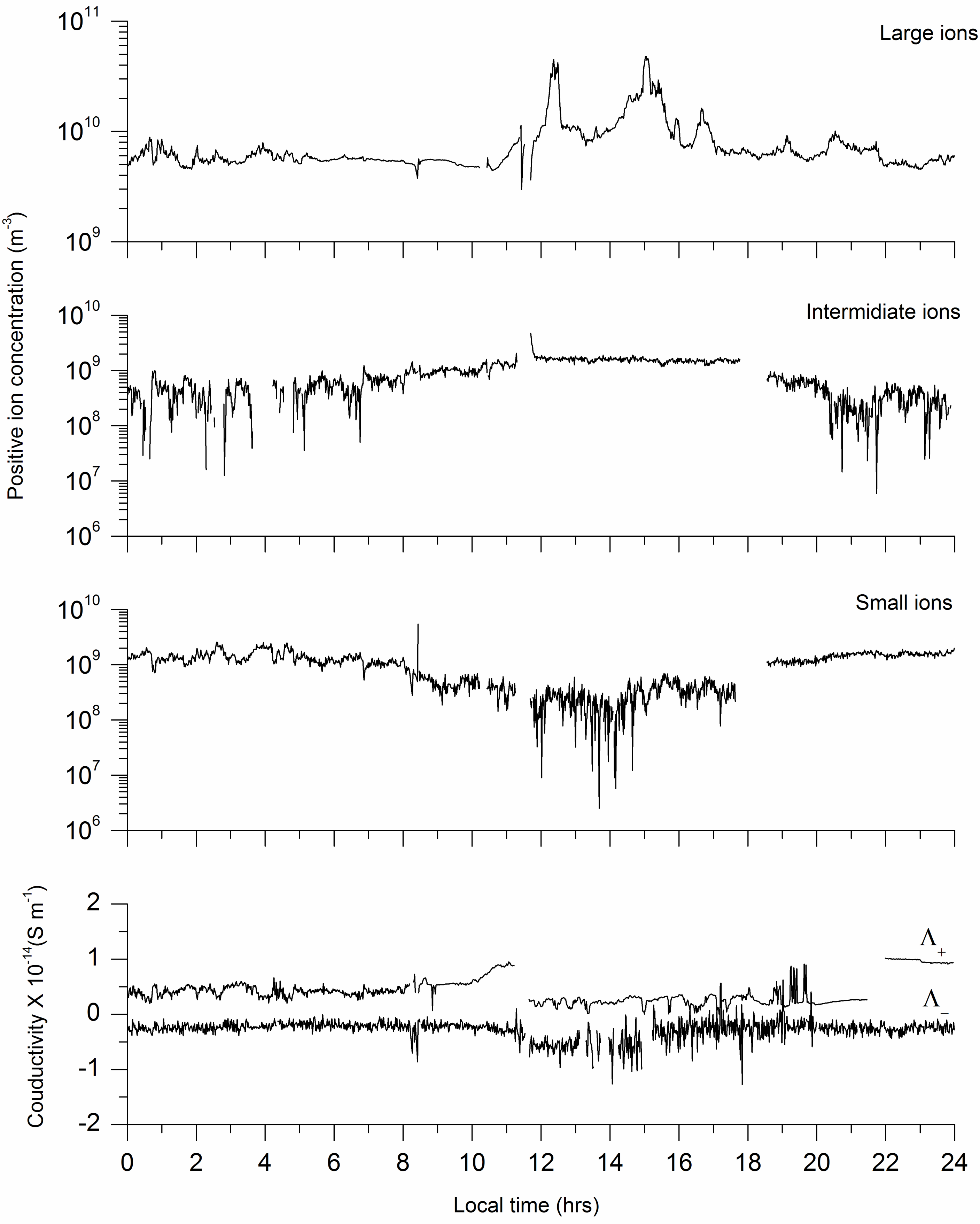

Fig. 4

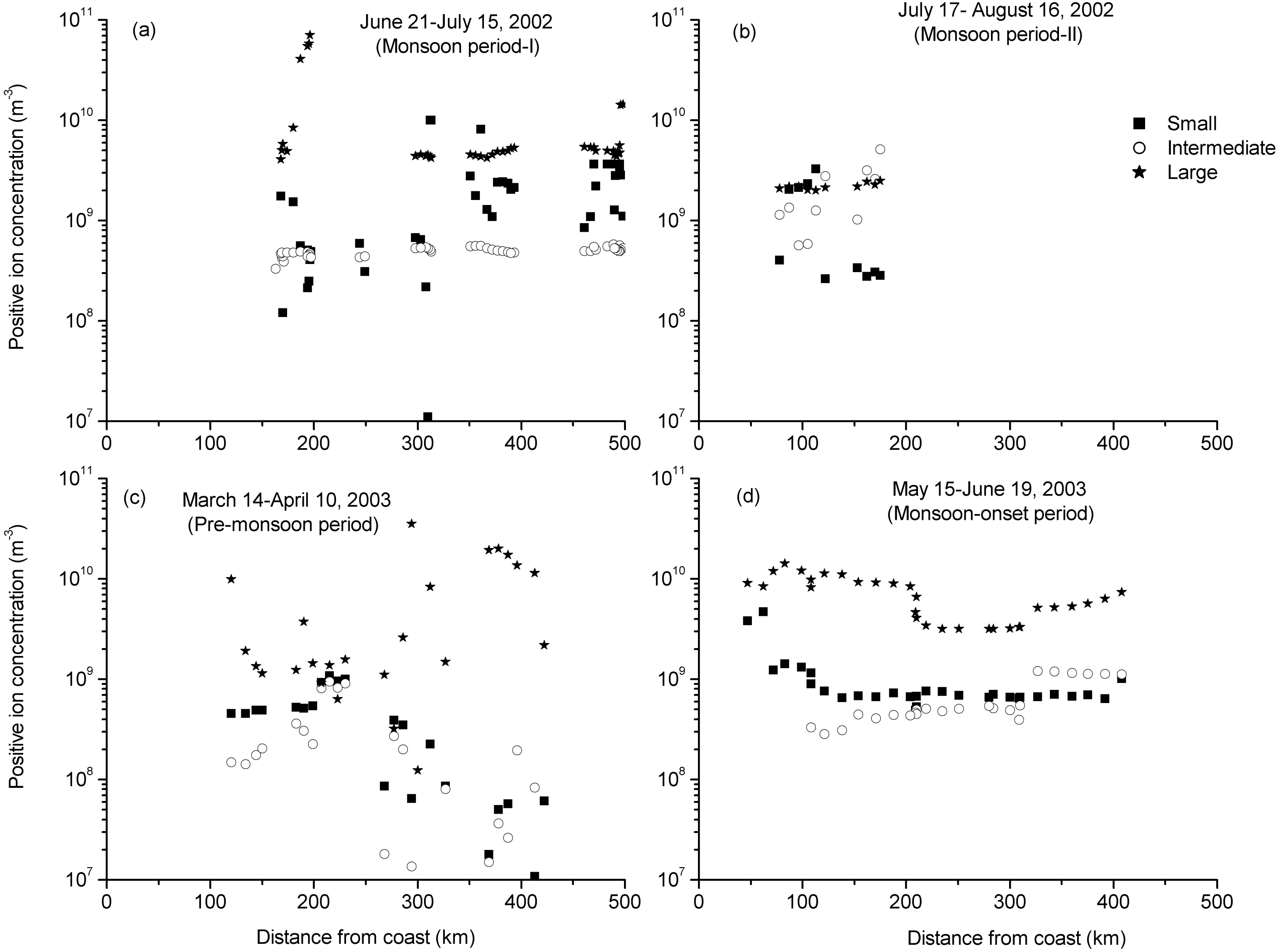

(a)
June 21-July 15, 2002
(Monsoon period-I)
(b)
July 17- August 16, 2002
(Monsoon period-II)
(c)
March 14-April 10, 2003
(Pre-monsoon period)
(d)
May 15-June 19, 2003
(Monsoon-onset period)
Small
Intermediate
Large
Positive ion concentration (m$^{-3}$)
Distance from coast (km)
$10^{7}$
$10^{8}$
$10^{9}$
$10^{10}$
$10^{11}$
0
100
200
300
400
500

Fig. 5

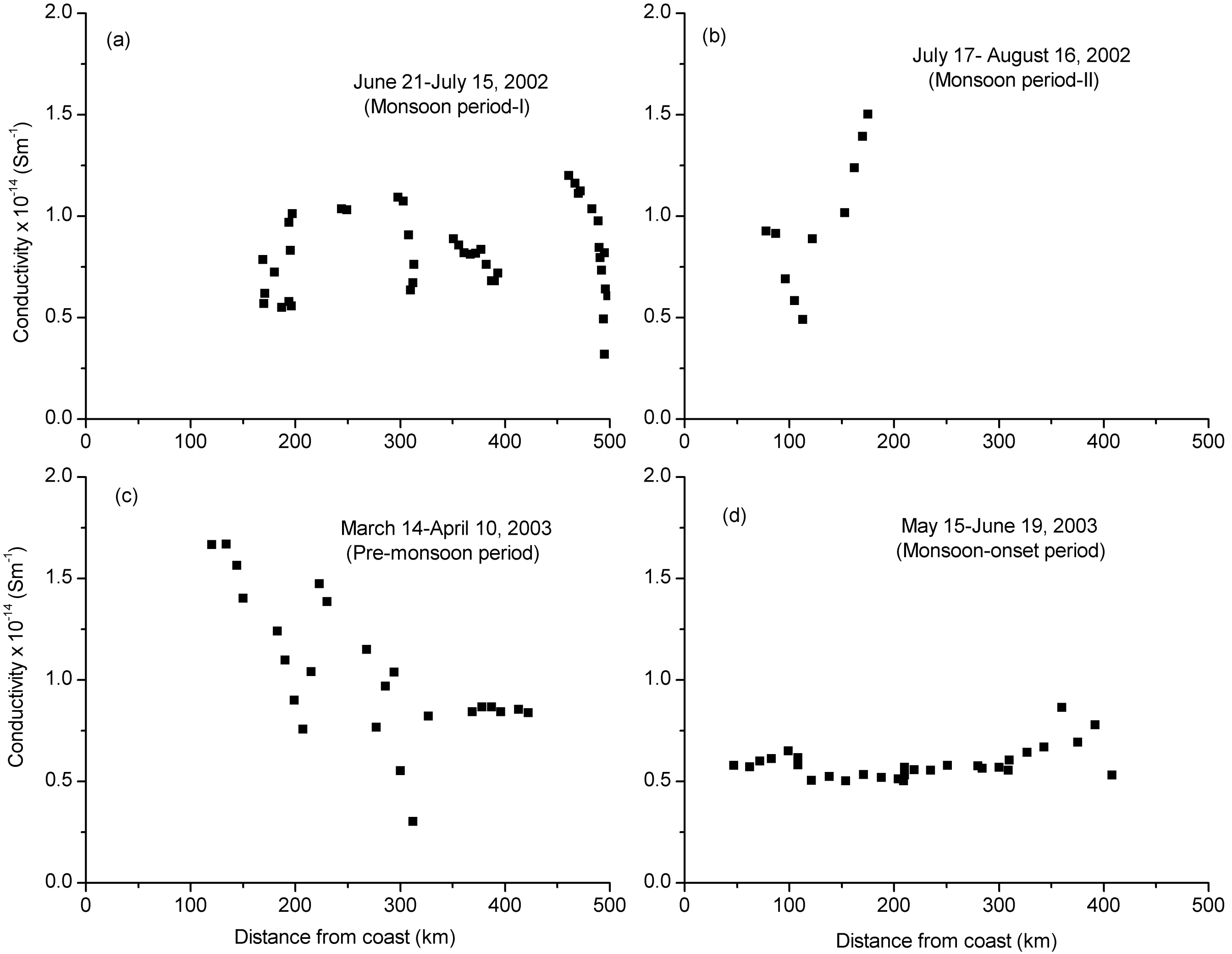
(a)
June 21-July 15, 2002
(Monsoon period-I)
(b)
July 17- August 16, 2002
(Monsoon period-II)
(c)
March 14-April 10, 2003
(Pre-monsoon period)
(d)
May 15-June 19, 2003
(Monsoon-onset period)
Conductivity x 10^-14 (Sm^-1)
Distance from coast (km)
0.0
0.5
1.0
1.5
2.0
0
100
200
300
400
500

Fig. 6

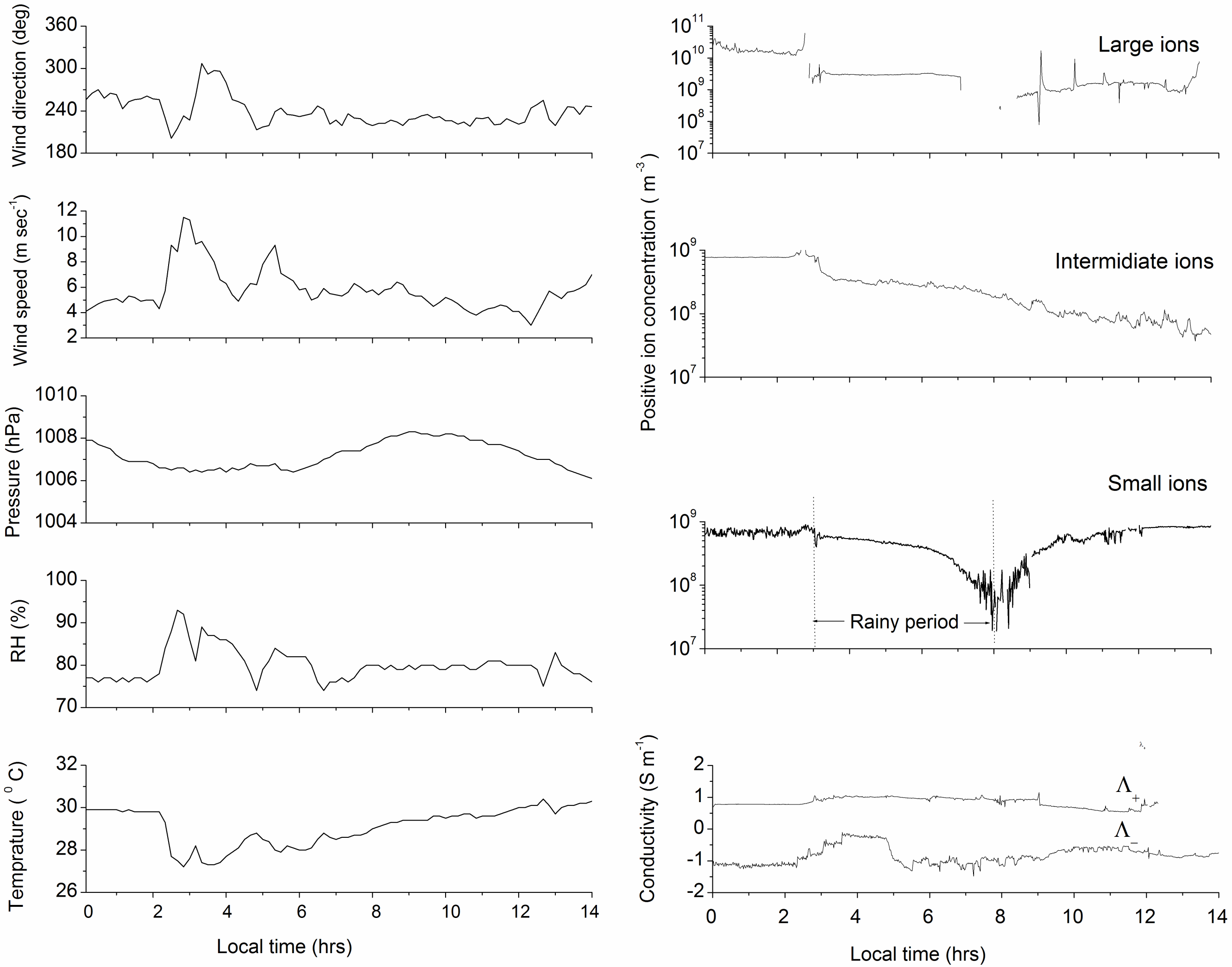

Fig.7

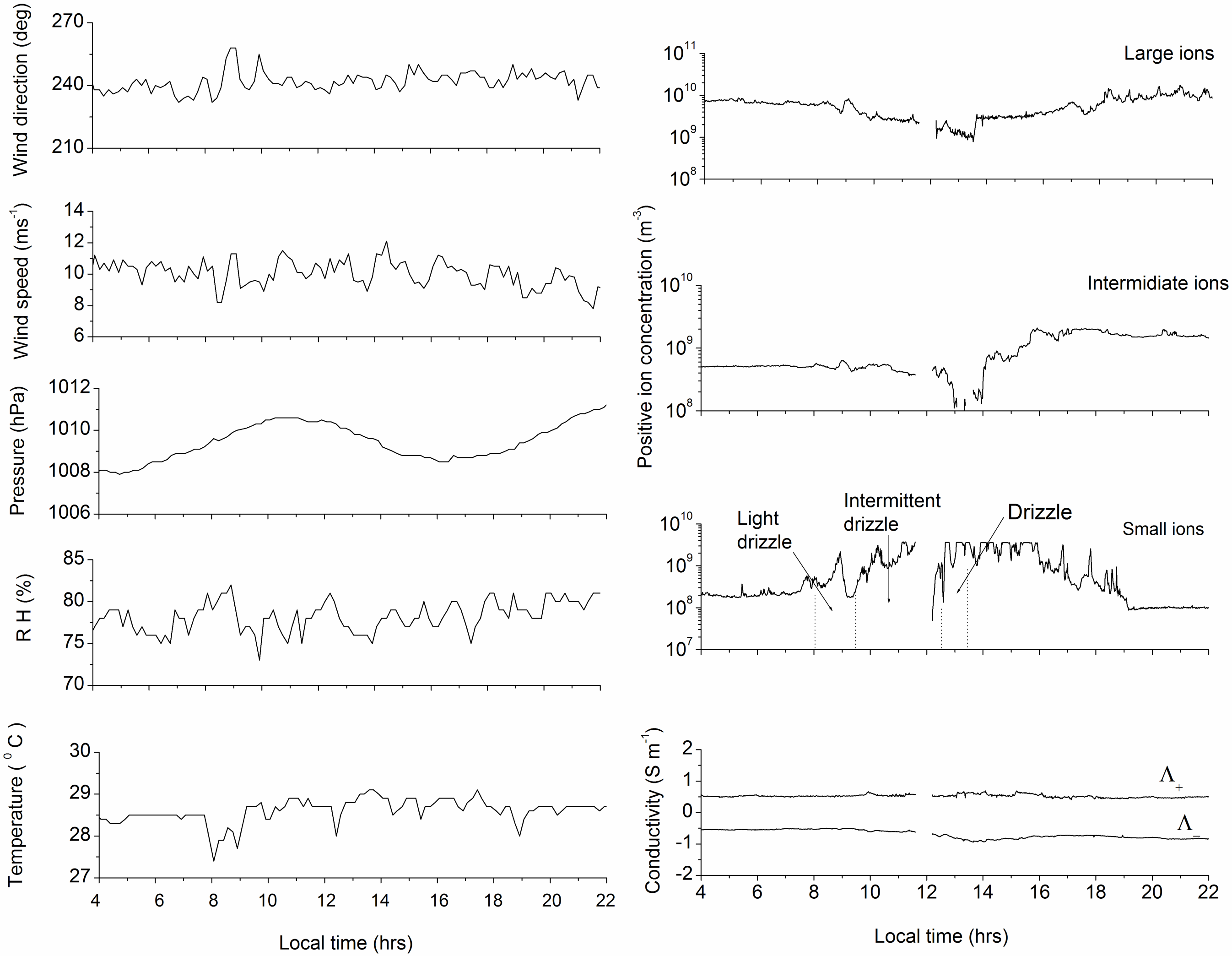

Fig 8

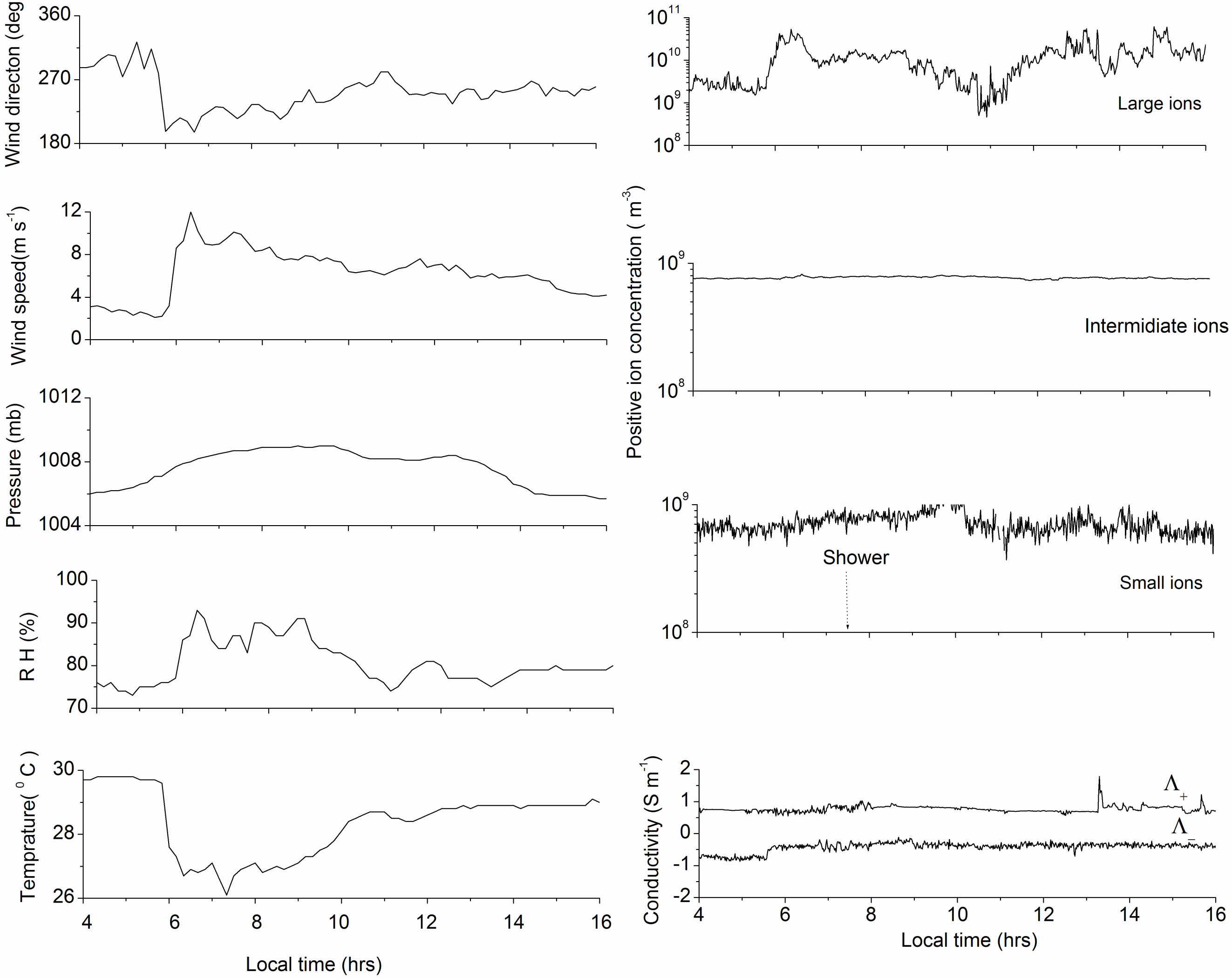